\documentclass[sigconf,nonacm]{acmart}
\AtBeginDocument{%
  }

\acmConference[JAW'26]{Evaluating RE Practices for Explainability: Synthesizing Insights from Daimler Truck into an Explainable RE Framework Proposal}{April 12--18,
  2026}{Rio De Janeiro, Brazil}

\begin{document}

\title{Evaluating RE Practices for Explainability: Synthesizing Insights from Daimler Truck into an Explainable RE Framework Proposal}

\author{Umm-e- Habiba}
\email{umme.habiba@tum.de}
\orcid{0000-0001-8953-9624}
\affiliation{%
  \institution{TUM School of CIT, Technical University of Munich}
  \city{Heilbronn}
  \country{Germany}
}

\author{Lucas Mauser}
\orcid{0000-0002-9235-5145}
\affiliation{%
  \institution{Daimler Truck AG, Leinfelden-Echterdingen}
  \country{Germany}}
\email{lucas.mauser@daimlertruck.com}

\author{Jonas Fritzsch}
\orcid{0000-0002-6121-2731}
\affiliation{%
  \institution{Institute of Software Engineering, University of Stuttgart}
  \city{Stuttgart}
  \country{Germany}}
  \email{jonas.fritzsch@iste.uni-stuttgart.de}

\author{Justus Bogner}
\orcid{0000-0001-5788-0991}
\affiliation{%
 \institution{Department of Computer Science, Vrije Universiteit Amsterdam}
 \city{Amsterdam}
 \country{Netherlands}}
 \email{j.bogner@vu.nl}

\author{Stefan Wagner}
\orcid{0000-0002-5256-8429}
\affiliation{%
 \institution{TUM School of CIT, Technical University of Munich}
  \city{Heilbronn}
  \country{Germany}}
  \email{stefan.wagner@tum.de}

\renewcommand{\shortauthors}{Habiba et al.}

\begin{abstract}
Explainability has emerged as a critical requirement for AI-based systems, particularly in safety-critical and regulated domains. Although prior research has proposed frameworks, patterns, and user-centered approaches to support explainability, there is limited empirical understanding of how existing Requirements Engineering (RE) practices support explainability requirements across the RE lifecycle, especially in an industrial context. This paper reports early findings from an ongoing industry-based study investigating how explainability requirements are elicited, specified, and validated using established RE techniques. We conducted a multi-phase qualitative study with eight practitioners at Daimler Truck, employing think-aloud protocols and moderated group discussions across requirements elicitation, specification, and validation steps. Our preliminary analysis reveals recurring challenges across all steps, including conceptual ambiguity during elicitation, limited testability and expressiveness during specification, and fragmented validation due to vague criteria and regulatory uncertainty. These findings indicate that current RE practices provide limited support to systematically address explainability requirements. The paper contributes empirical insights into step-specific and cross-cutting challenges and outlines a research vision toward developing an empirically grounded RE framework for explainable AI-based systems.
\end{abstract}


\keywords{Explainable AI, Requirements Engineering, Framework Proposal, User Study, Focus Group}


\maketitle

\section{Introduction}
Artificial intelligence (AI) is increasingly used in domains such as medicine~\cite{caruana2015intelligible,jin2020artificial}, law~\cite{jin2023invisible}, autonomous driving~\cite{atakishiyev2021explainable}, and loan application approval~\cite{sachan2020explainable}. In these and other high-stakes settings, AI-based systems often support or influence critical decision-making processes. Consequently, responsible users, including physicians, judges, drivers, and bankers, require varying levels of explanation to appropriately trust, assess, and act on AI outputs. Ensuring that AI decisions are interpretable is essential for safe, responsible, and legally compliant AI deployment~\cite{longo2020explainable,lagioia2020impact}.

From a software engineering perspective, developing explainable AI-based systems requires effective methodologies to elicit, specify, and validate explainability-related requirements. However, as explainability has only recently emerged as a key non-functional requirement~\cite{kohl2019explainability}, practitioners currently receive limited support for systematically integrating explainability into Requirements Engineering (RE) processes. Explainability poses particular challenges for RE due to its context-dependent, multi-dimensional, and human-centered nature.

Prior research has increasingly investigated explainability from an RE perspective by proposing conceptual frameworks, experience reports, and user-centered methods that support specific explainability-related activities~\cite{11030046,taylor2025eliciting,11135747}. In parallel, several studies have explored explainability requirements in real-world contexts through interactive and contextual user studies, focusing on particular stakeholders, domains, or techniques~\cite{sporsem2025clinicians,10260835,deters2023means,9920064}. However, existing work predominantly addresses isolateed steps or techniques within the RE process. Most frameworks focus on a single RE step, such as elicitation, specification, or validation, while real-world studies typically investigate individual practices or methods in isolation. Consequently, current literature does not provide a holistic understanding of how explainability requirements are engineered end-to-end, nor how different RE practices interact across the RE lifecycle in practical settings.

To the best of our knowledge, no prior study has systematically examined the complete RE process for explainability from elicitation through specification to validation in a real-world setting. Moreover, existing studies do not compare multiple techniques within each RE step, nor do they analyze the challenges and trade-offs that arise when these techniques are applied sequentially as part of an integrated RE process.
To address this gap, we report early findings from an ongoing industry-based study that examines how practitioners apply established RE practices to explainability requirements in AI-based systems. As a first step toward forming a dedicated explainability-focused RE framework, we assessed the effectiveness of existing RE techniques in the elicitation, specification, and validation steps. We conducted a multi-phase user study with Daimler Truck AG, one of the world’s largest commercial vehicle manufacturers~\cite{DTAG}. In the first phase, eight practitioners applied established RE practices to a real-world scenario drawn from their organizational context while following a think-aloud protocol. This approach allowed us to capture the challenges, rationales for decision-making, and underlying reasoning as explainability requirements were handled in practice. The insights derived from this phase, together with evidence from our previous research~\cite{habiba2024mature,habiba2025ml}, inform our ongoing effort to improve RE support for explainability.
The contributions of this paper are as follows:
\begin{enumerate}
    \item We provide empirical insights into how existing RE practices perform when applied to explainability requirements in industrial AI-based systems.
    \item We identify and synthesize key challenges, limitations, and gaps that arise during the elicitation, specification, and validation of explainability requirements in practice.
    \item We outline a research vision toward developing an empirically grounded RE framework to better support the systematic handling of explainability requirements.
\end{enumerate}
\section{Study Design}
We structured our study into three phases. In the first phase, we examined how effectively existing RE practices address explainability challenges in AI-based systems. To assess the application of RE techniques across elicitation, specification, and validation, we employed think-aloud protocols~\cite{ericsson1980verbal}.

\begin{table*}[ht!]
\fontsize{8.5}{10}\selectfont
\centering
\caption{Participant Demographics}
\label{tab:participants_demographics}

\renewcommand{\arraystretch}{1.1} 
\resizebox{\textwidth}{!}{%
\begin{tabular}{
>{\centering\arraybackslash}m{0.9cm}
>{\raggedright\arraybackslash}m{2.1cm}
>{\raggedright\arraybackslash}m{1.6cm}
>{\raggedleft\arraybackslash}m{1.4cm}
>{\centering\arraybackslash}m{1.4cm}
>{\centering\arraybackslash}m{1.4cm}
>{\raggedright\arraybackslash}m{1.9cm}
>{\raggedright\arraybackslash}m{2.3cm}
>{\raggedright\arraybackslash}m{1.8cm}}
\toprule
\textbf{P-ID$^a$} & \textbf{Background$^b$} & \textbf{Role} & \textbf{Experience} &
\textbf{Experience$^c$} & \textbf{Experience$^c$} &
\textbf{Elicitation} & \textbf{Specification} & \textbf{Validation} \\
 &  &  &  & \textbf{with RE} & \textbf{with AI} &  &  &  \\
\midrule

P1 & Automotive E/E Architecture & Development Engineer & 1--3 years & Basic & Basic &
Task Analysis & Structured Natural Language & Walkthroughs \\

P2 & Electronic Architecture & Product Engineer & 7--10 years & Moderate & Basic &
Task Analysis & Use Case Modelling & Walkthroughs \\

P3 & ADAS & Product Engineer & 7--10 years & Moderate & Moderate &
Interview & Structured Natural Language & Compliance Check \\

P4 & Artificial Intelligence & Researcher & 1--3 years & Moderate & Moderate &
Interview & Use Case Modelling & Compliance Check \\

P5 & Active Safety & Safety Engineer & 7--10 years & Moderate & Moderate &
Brainstorming & Softgoal Interdependency & Walkthroughs \\

P6 & Autonomous Driving & Researcher & 1--3 years & Moderate & Moderate &
Brainstorming & Use Case Modelling & Structured Feedback \\

P7 & Automotive E/E Architecture & Development Engineer & $>$10 years & Advanced & Moderate &
Brainstorming & Structured Natural Language & Structured Feedback \\

P8 & Automotive E/E Architecture & Development Engineer & 4--6 years & Moderate & Moderate &
Interview & Softgoal Interdependency & Structured Feedback \\

\bottomrule
\end{tabular}%
}

\begin{flushleft}
\fontsize{6.5}{10}\selectfont
\vspace{1mm}
\hspace{4mm} $^a$ P-ID = Participant ID $^b$ E/E (Electrical/Electronic) $^b$ ADAS = Advanced Driver Assistance Systems
$^c$ Experience: Basic = 1--3 years, Moderate = 4--6 years, Advanced = 7 years or more
\end{flushleft}

\end{table*}

In the second phase, we will systematically analyse the qualitative data collected in the first phase to derive empirical insights that inform the design of a dedicated RE framework for explainability. In the third phase, we plan to evaluate this framework with practitioners through focus groups~\cite{krueger2002designing}, aiming to capture practitioners’ experiences and identify further improvement opportunities. In the following sections, we describe the study design for the first phase only, as the second and third phases are still ongoing and will not be discussed here.

\subsection{Participant Sampling and Demographics}
To guide participant selection, we applied the following criteria:
\begin{itemize}
    \item At least two years of experience working on AI/ML projects.
    \item Familiarity with RE processes and steps.
\end{itemize}

\noindent
As shown in Table~\ref{tab:participants_demographics}, participants were recruited from Daimler Truck through established industry connections within the research team. To ensure practical relevance, we used a real-world industrial scenario from Daimler Truck AG based on the Active Side Guard Assist (ASGA) feature~\cite{ASGA}, mandated under the General Safety Regulation (GSR) 2019/2144~\cite{GSR1} and UN Regulation No.~151~\cite{BSIS}. Participants’ familiarity with the system and organizational context enabled a more realistic assessment of limitations in current RE practices.
Below, we explain Phase One. Phases Two and Three are currently in progress; therefore, they are not discussed or described at this stage.
\subsection{Phase One: Think-Aloud Protocol}
We employed a think-aloud protocol to observe how practitioners apply existing RE elicitation, specification, and validation techniques when handling explainability requirements in real time~\cite{ericsson1998study}. This approach allowed us to uncover context-specific challenges that are often difficult to capture through retrospective or literature-based methods.

\subsubsection{Design}
We focus on elicitation, specification, and validation as the core steps of the RE lifecycle~\cite{nuseibeh2000roadmap,sommerville2016software}. Explainability introduces distinct challenges at each step: stakeholders often struggle to articulate explainability needs during elicitation~\cite{suresh2021}, specification is complicated by the context-dependent and multi-dimensional nature of explainability~\cite{amershi2019guidelines}, and validation requires assessing human-centered interpretability rather than purely technical correctness~\cite{doshi2017towards}. Examining these steps jointly enables a holistic and methodologically grounded assessment of how well existing RE practices support explainability requirements in AI-based systems.

Accordingly, we selected three established RE practices per step (see Table~\ref{tab:consolidated_challenges}) to empirically evaluate their effectiveness. These techniques were chosen based on prior work identifying practical gaps in RE for AI-based systems~\cite{habiba2024mature,balasubramaniam2024candidate} and their ability to address recurring challenges such as ambiguous abstraction levels, cross-cutting dependencies, and trade-off representation~\cite{habiba2025ml}.

\subsubsection{Execution and Data Analysis}
Eight participants were divided into three groups, each assigned a different elicitation technique. During the first session, groups independently applied their assigned techniques using a think-aloud protocol; challenges were documented on structured sheets and all sessions were audio-recorded. The elicited requirements were subsequently consolidated into a unified document by the researchers and participants to ensure consistency and remove redundancy.

This document served as input to the specification step, where participants were reshuffled to reduce team-specific bias and applied different specification techniques. Following both the elicitation and specification steps, group discussions were conducted to reflect on findings and identify challenges. The finalized requirements document was then used in the validation step, during which participants applied various validation techniques while verbalizing encountered challenges, followed by a final collective discussion.

All audio recordings were transcribed and analyzed using the constant comparison method~\cite{seaman2008qualitative}, guided by grounded theory principles. An initial coding scheme was iteratively refined through collaborative analysis by three authors, who repeatedly revisited the data, discussed interpretations, and merged or split codes as the analysis progressed
\noindent
\section{Preliminary Results}
 \label{sec:Rq1}
 \smallskip

\noindent
This section reports preliminary findings from the first phase (Think-Aloud Protocol) of our study. The results are based on observations from think-aloud sessions and subsequent group discussions. During the group discussions, all participants collectively reflected on their experiences across the three think-aloud sessions. Table~\ref{tab:consolidated_challenges} summarizes the preliminary challenges identified across elicitation, specification, and validation steps. While the nature of these challenges varied by RE steps and technique, several cross-cutting patterns emerged, suggesting systemic limitations in current RE practices rather than isolated methodological shortcomings.\\

\begin{table}[t]
\renewcommand{\arraystretch}{0.9}
\fontsize{8}{10}\selectfont
\caption{Preliminary Challenges for Explainability \\
Requirements Across RE Steps}
\label{tab:consolidated_challenges}

\centering
\begin{tabular}{p{2.5cm} p{5.4cm}}
\toprule
\textbf{RE Practice} & \textbf{Key Challenges} \\
\midrule
\textbf{Elicitation} \\
\midrule

\textbf{Interviews} &
Divergent stakeholder understanding; abstract explainability concepts; lack of shared terminology. \\

\textbf{Brainstorming} &
Explainability deprioritized; unstructured discussion; dominance of few participants. \\

\textbf{Task Analysis} &
Explainability needs hard to map to workflows; weak goal-function linkage. \\

\midrule
\textbf{Specification} \\
\midrule
\textbf{Structured NL} &
Ambiguous abstraction levels; overlapping requirements; limited testability. \\

\textbf{Use Case Diagrams} &
Poor support for cross-cutting explainability concerns; missing acceptance criteria. \\

\textbf{Softgoal Graphs} &
Complex interdependencies; high cognitive load; early system-level reasoning required. \\

\midrule
\textbf{Validation} \\
\midrule
\textbf{User Feedback} &
Vague, non-measurable requirements; context-dependent assessment difficulties. \\

\textbf{Walkthroughs} &
Underspecified requirements; limited traceability and stakeholder coverage. \\

\textbf{Compliance Checking} &
Ambiguous legal terms; transparency confidentiality tension; weak regulatory traceability. \\

\bottomrule
\end{tabular}
\end{table}

\noindent
\subsection{Challenges in the Elicitation Step}

During elicitation, participants consistently struggled to articulate explainability requirements using interviews, brainstorming, and task analysis techniques. A recurring issue was the absence of a shared understanding of what constitutes an “explanation.” Participants interpreted explainability differently depending on their background and role, leading to vague, abstract, or conflicting requirements.
Additionally, explainability was often deprioritized relative to more concrete and measurable requirements such as performance or safety. This tendency was particularly evident during brainstorming sessions, where discussions gravitated toward familiar system properties, leaving explainability insufficiently explored. For task analysis, participants found it difficult to identify where explainability should be integrated into workflows, highlighting a gap between abstract explainability goals and concrete system functions.
Together, these observations suggest that existing elicitation techniques provide limited support for externalizing and structuring explainability needs, especially when stakeholders lack established mental models or vocabulary for explainability

\subsection{Challenges in the Specification Step}

In the specification step, participants applied Structured Natural Language (SNL), Use Case Diagrams (UCD), and Softgoal Interdependency Graphs (SIG) to document explainability requirements. Across all techniques, participants encountered difficulties in translating elicited explainability needs into precise, testable specifications.
With SNL, explainability requirements frequently overlapped with other non-functional requirements such as transparency, fairness, and accountability. These requirements were expressed at inconsistent levels of abstraction. Although SNL encouraged detailed descriptions, participants reported uncertainty about how to make explainability requirements measurable or verifiable. UCDs, in contrast, offered limited expressive power for capturing explainability as a cross-cutting concern, resulting in omissions and oversimplifications. SIGs enabled participants to reason about trade-offs and dependencies involving explainability but were perceived as complex and difficult to manage without a stable system understanding.
These findings indicate that current specification techniques inadequately support the representation of context-dependent, evolving, and human-centered explainability requirements, often forcing 

\smallskip\noindent
\subsection{Challenges in the Validation Step}
The validation step revealed further limitations when assessing whether explainability requirements had been satisfactorily addressed. Participants reported that vague or underspecified requirements made it difficult to evaluate explainability using Structured User Feedback and Walkthroughs. In many cases, validation discussions focused on linguistic clarity rather than whether explanations effectively supported user understanding or decision-making.
Compliance Checking was the only technique that explicitly raised regulatory considerations; however, participants struggled to interpret abstract legal terminology and to reconcile legal expectations with technical feasibility. The absence of legal experts and measurable acceptance criteria further constrained meaningful validation.
Overall, validation challenges were not solely attributable to the selected techniques but also reflected upstream issues originating in elicitation and specification. Ambiguities introduced early in the RE process propagated into validation, limiting traceability, accountability, and confidence in explainability outcomes.
\vspace{-1mm}
\subsection{Cross-Step Synthesis}

Across all steps, the preliminary findings suggest that challenges in handling explainability requirements are interdependent and cumulative. Conceptual ambiguity during elicitation leads to under-specified requirements, which in turn hinder meaningful validation. Particularly regarding what constitutes an adequate explanation? These patterns indicate that explainability is insufficiently supported as a first-class concern in existing RE practices and is instead treated in an ad-hoc manner across phases.
Rather than pointing to the failure of individual techniques, these results highlight a need for better process-level guidance that helps practitioners maintain continuity, traceability, and shared understanding of explainability throughout the RE lifecycle.
\section{Research Vision}

\noindent
Our vision emerging from this study is to support explainability as a systematically engineered concern throughout the RE lifecycle, rather than as an isolated artifact or post-hoc design consideration. Our preliminary findings suggest that current challenges in explainability engineering stem from a lack of awareness among practitioners and also from limitations in how existing RE practices structure, refine, and validate explainability requirements.

In particular, the observed breakdowns across elicitation, specification, and validation indicate a need for process-oriented support that explicitly addresses the evolving, context-dependent, and human-centered nature of explainability. Such support should help practitioners articulate explainability needs early, maintain coherence across RE steps, and assess explainability using meaningful and context-sensitive criteria.

\textbf{In the second phase}, our goal is to develop an empirically grounded RE framework that builds on observed practices and challenges to enhance the systematic handling of explainability requirements. Rather than replacing existing RE techniques, the framework will augment them by making trade-offs explicit and improving traceability across development phases. It will also support alignment between stakeholder expectations, technical feasibility, and regulatory considerations. \textbf{In phase three}, through continued empirical investigation and feedback from the same Daimler Truck practitioners, we aim to refine this framework and contribute actionable guidance for engineering explainability in AI-based systems.
\vspace{-6mm}
\section{Conclusion and Future Work}
We present early findings from an ongoing industry-based study examining how explainability requirements are elicited, specified, and validated using established RE practices. The primary challenge observed during elicitation was a lack of shared understanding of explainability itself, as participants held differing interpretations of what explainability should provide and for whom. This initial conceptual misalignment propagated across subsequent RE steps, resulting in limited expressiveness and testability during specification, and fragmented validation due to unclear criteria and regulatory uncertainty. Overall, these findings indicate that current RE practices offer limited support for treating explainability as a first-class, end-to-end concern throughout the RE lifecycle.

As an exploratory study, the presented results are preliminary and intended to provide first empirical insights rather than definitive conclusions. Nevertheless, they reveal cross-phase patterns suggesting that challenges in handling explainability requirements are cumulative and process-related, rather than isolated to individual techniques. This underscores the need for holistic, process-oriented guidance that supports continuity, traceability, and shared understanding of explainability across RE process.

As the next phase, we will develop an empirically grounded RE framework that builds on the identified challenges to better support the systematic engineering of explainability requirements in AI-based systems. This framework will be refined and evaluated in collaboration with industry practitioners through focus groups. In addition, we plan to extend the study to other organizational and domain contexts to assess the transferability of the findings and to further strengthen empirical grounding.

\bibliographystyle{ACM-Reference-Format}
\bibliography{references}

\end{document}